\documentclass[intlimits,twoside,a4paper]{article}

\usepackage{amsmath,amssymb}
\usepackage{graphicx}
\usepackage[cp1251]{inputenc}

\usepackage{cmpj3}

\issue{2017}{20}{3}{33803}
\doinumber{10.5488/CMP.20.33803}

\title[Porosity of a passive layer]
{Effects of porosity in a model of corrosion and passive layer growth}
\author[F.D.A. Aar\~ao Reis]{F.D.A. Aar\~ao Reis\thanks{E-mail: reis@if.uff.br}}
\address{
Instituto de F\'\i sica, Universidade Federal Fluminense,
Avenida Litor\^anea s/n, 24210-340 Niter\'oi RJ, Brazil
}

\date{Received April 24, 2017, in final form June 23, 2017}
\begin{document}
\maketitle

\begin{abstract}
We introduce a stochastic lattice model to investigate the effects of pore formation in
a passive layer grown with products of metal corrosion.
It considers that an anionic species diffuses across that layer and reacts at the corrosion
front (metal-oxide interface), producing a random distribution of compact regions and
large pores, respectively represented by O (oxide) and P (pore) sites.
O sites are assumed to have very small pores, so that the fraction $\Phi$ of P sites
is an estimate of the porosity, and the ratio between anion diffusion coefficients in those
regions is $D_{\text r}<1$.
Simulation results without the large pores ($\Phi =0$) are similar to those of a formerly
studied model of corrosion and passivation and are explained by a scaling approach.
If $\Phi >0$ and $D_{\text r}\ll 1$, significant changes are observed in passive layer growth
and corrosion front roughness.
For small $\Phi$, a slowdown of the growth rate is observed, which is interpreted as a
consequence of the confinement of anions in isolated pores for long times.
However, the presence of large pores near the corrosion front increases the frequency of
reactions at those regions, which leads to an increase in the roughness of that front.
This model may be a first step to represent defects in a passive layer which favor
pitting corrosion.


\pacs 82.45.Bb, 05.40.-a, 68.35.Ct

\end{abstract}

\section{Introduction}

\begin{figure}[!t]
\center
\includegraphics[clip,width=0.9\textwidth, angle=0]{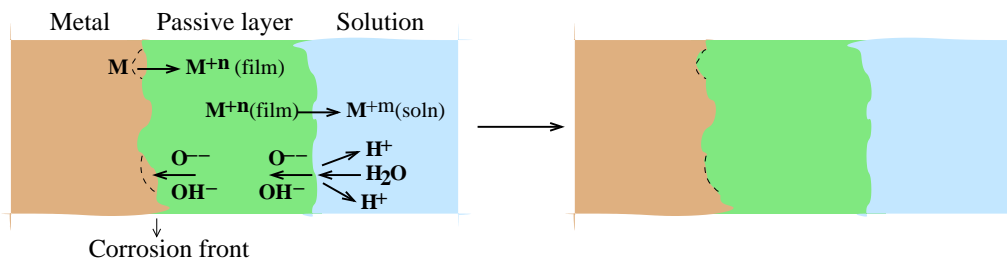}
\caption{(Color online) Scheme of the phases in the corrosion process and transport of ions in the passive
layer. The advance of the corrosion front occurs where those ions react, leading to a growth
of the passive layer at the regions indicated with dashed curves.
}
\label{scheme}
\end{figure}

If a metal is in contact with an aggressive solution, a corrosion process begins and
leads to the formation of a layer of oxide or hydroxide which protects the metal.
However, the corrosion process continues due to the transport of ions across that
passive layer.
Figure~\ref{scheme} illustrates this process.
In conditions where dissolution of the passive layer is very slow, its thickness increases
in time and the corrosion rate decreases.
Several models were already proposed for this process and were successfully applied
to describe experiments with various materials
\cite{chao,macdonald2001,olsson2003,zhang}.
Stochastic models in lattices (also called cellular automata or kinetic Monte Carlo)
were also studied by several authors to represent coarse grained features of passive layers,
such as temporal and spatial scaling of the thickness and of the roughness of the interfaces
with the metal (corrosion front; see figure~\ref{scheme}) and the solution
\cite{saunier2004,taleb2004,saunier2005,incubation,passive1,passive2,dung2010,rita2000}.

Reference \protect\cite{passive2} proposed a model in which diffusion of anions was the main
transport mechanism and in which reactions occurred at the corrosion front.
This model implicitly assumed that the passive layer had pores where ion transport
is possible, but that layer was considered homogeneous.
A scaling approach predicted a crossover between an initial regime with constant velocity
of the corrosion front and rapid roughening (Kardar-Parisi-Zhang (KPZ) scaling \cite{kpz})
and a long time regime with parabolic ($t^{1/2}$) displacement and slow roughening
(diffusion-limited erosion (DLE) scaling \cite{krugmeakin}).
Recently, reference~\cite{alteredlayer} extended that model to represent the
dissolution-reprecipitation mechanism \cite{hellmann2012} that leads to the formation of an
altered layer during the weathering of a mineral.

In this work, we introduce a corrosion-passivation model similar to that of
 \protect\cite{passive2} but considering an inhomogeneous passive layer, with random
distribution of compact regions and large pores.
Such pores may be filled with solution, act as traps for the moving anions, and eventually
form preferential paths for diffusion.
On the other hand, diffusion in the remaining parts of the layer (called the compact region)
is assumed to be much slower.
Simulations of the model in two dimensions (square lattice) show that
the thickening of the passive layer and the roughning of the corrosion front
depend on the porosity and on the ratio of diffusion
coefficients in the compact regions and in the large pores.
For small values of that ratio, a remarkable decrease in the growth rate may be observed
for small porosity, while the roughness of the corrosion front increases.

The defects of the oxide or hydroxide layer formed during metal corrosion play an important
role in pitting corrosion \cite{maurice2012}; for instance, they facilitate the
transport of aggressive anions from the electrolyte to the metal surface and increase the
frequency of metastable pitting events \cite{amin2014}.
This model of an inhomogeneous passive layer is a first step to represent such defects and
determine their possible effects.

The rest of this paper is organized as follows.
In section~\ref{sectionmodel}, we present the model and briefly discuss the relations with other models
of passive layer growth.
In section~\ref{sectiongrowth}, we analyze the time evolution of the thickness of the passive
layer.
In section~\ref{sectionroughness}, we analyze the time evolution of the roughness of the
corrosion front.
In section~\ref{conclusion}, we summarize our results and present our conclusions.

\section{The model of corrosion and passive film growth}
\label{sectionmodel}

\begin{figure}[!t]
\center
\includegraphics[clip,width=0.9\textwidth, angle=0]{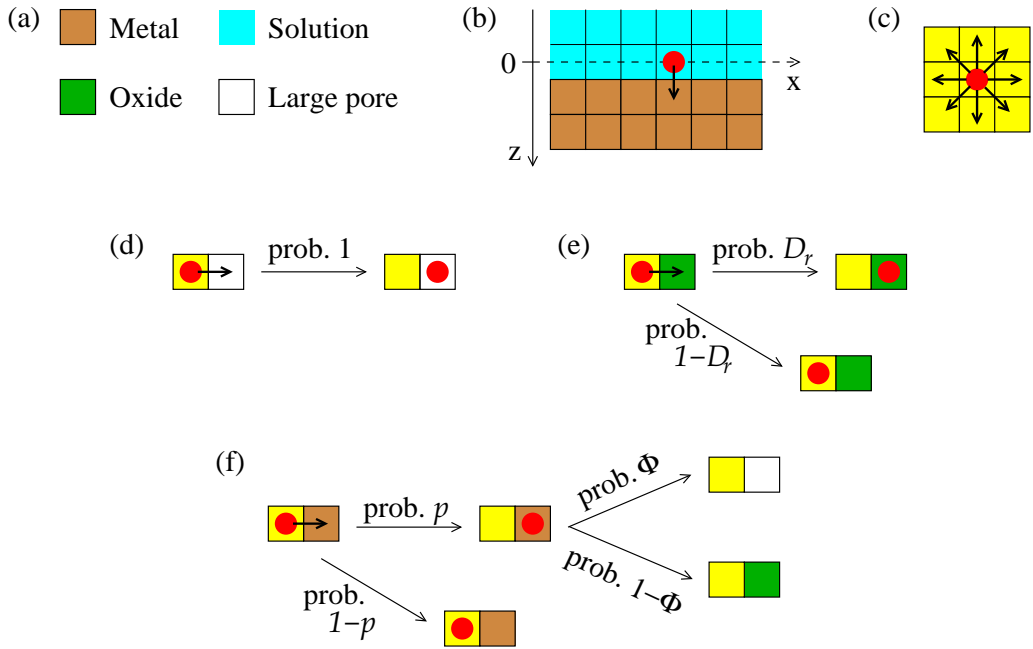}
\caption{(Color online) (a) Four types of the lattice sites.
(b) Initial configuration of the lattice with a particle A (red circle) at its initial position
in the line $z=0$.
(c) Possible directions of step attempts of the particle A. Yellow color is used here
for sites that may be of any type.
(d)--(f) Step attempts of a particle A with target P, O, and M sites, respectively.
The step probability does not depend on the current site (yellow) of the particle A.
}
\label{model}
\end{figure}

The model is defined in a square lattice and the lattice constant is taken as the length unit.
Each site is expected to represent a homogeneous region with several atoms or molecules.
Thus, the lattice constant corresponds to some nanometers or larger sizes in possible applications.
The sites may be of four types, as shown in figure~\ref{model}~(a):
metal (M), oxide (O), solution (S) at $z\leqslant 0$, and pore (P) at $z> 0$.

The configuration at time $t=0$ has solution at all
points with $z\leqslant 0$ and metal at all points at $z> 0$, as shown in figure~\ref{model}~(b).
The passive layer is produced at $z>0$ by continuous
transformation of M sites into O or P sites.
O sites represent regions with very small pores, which permit slow ion transport but which
have a negligible contribution to the total porosity; the set of O sites is called
the compact region of the layer.
P sites represent larger pores and their volume fraction is assumed to be equal to the
total porosity.

We assume that an anionic species is produced in electrochemical water decomposition
at the solid-solution interface $z=0$.
This species is represented by a mobile particle A that can occupy any lattice site;
one particle A is shown in the initial configuration of figure~\ref{model}~(b).
At each step of the corrosion-passivation process, one particle A is released at a randomly
chosen S site with $z=0$, executes a random walk on sites O and P, and eventually reacts
with a site M to produce a new O or P site.
A new particle~A is released only after the previous one has reacted; thus, a single
particle A is present in the lattice at each time.

The random walk of particle A considers a Moore neighbourhood \cite{gray}, in which
nearest neighbor (NN) and next nearest neighbor (NNN) sites are randomly chosen for each
step, as illustrated in figure~\ref{model}~(c) (yellow sites
are used to represent any type of lattice site in that figure).
The chosen neighbor is called the target site.
Possible outcomes of the step attempt are illustrated in figures~\ref{model}~(d)--(f).
If the target is an S site, then A does not move because the model does not
describe transport in solution (for this reason, such process is not shown in figure~\ref{model}).
If the target site is P [figure~\ref{model}~(d)], then the step is executed with probability $1$.
If the target site is O [figure~\ref{model}~(e)], then the step is executed with probability $D_{\text r}$;
we consider $D_{\text r}<1$, which represents a slower diffusion in the compact regions in comparison
with the diffusion in the large pores.
If the target site is M [figure~\ref{model}~(f)], then a reaction occurs with probability $p$;
the overall consequence of the reaction is represented by the annihilation of particle A and
the transformation of the M site into an O site (with probability $1-\Phi$) or into a P site
(with probability $\Phi$).

The stochastic rules of the model lead to the formation of a passive layer with a fraction
$\Phi$ of P sites.
This is the value of porosity because we assumed that pores of regions represented by
the O sites are very small.
The mechanism of pore formation considered here is an oversimplified description of a much more
complex process.
In a real corrosion-passivation process, we expect an increase in the volume of the oxide or
hydroxide film after reactions, which generates stress that may lead to rearrangement of
molecules.
However, the aim of our model is not to represent such details, but to investigate the effect
of inhomogeneity in the ion transport on the layer growth and on the morphology of the
corrosion front.

The parameter $D_{\text r}$ is a simple form of representing that inhomogeneity and
may be interpreted as a ratio of diffusion coefficients.
However, it is certainly a difficult quantity to be measured experimentally because it would
be necessary to measure two diffusion coefficients in very small homogeneous regions of a
passive layer.

We performed simulations of the model for several values of the parameters $p$, $D_{\text r}$,
and $\Phi$ using lattices of lateral size $L=512$ sites.
Average quantities were obtained considering $100$ different realizations of the
corrosion-passivation process for each parameter set; this is necessary to reduce the noise
effects, since a single realization does not represent all the relevant microscopic
environments.
The maximal thickness of the passive layer studied here is $H_{\text{max}}=400$, although for some
parameter sets the simulations were restricted to $H_{\text{max}}=100$.
The simulations were run in a dual Xeon E5520 computer with Linux for approximately six days.

In the implementation of our model, a new particle A is released at $z=0$ only after the previous
particle A has reacted.
This implies that the calculation of the time of layer growth is not as direct as in the
model of  \protect\cite{passive2}, in which the time scale was set by the
rates of the diffusion and reaction processes.
Here, we separate particles A in sets of $L$ particles that are consecutively released
(an average of one particle per column $x$) and assume that all particles of the $i$-th
set left their initial positions at time $i\delta t$, with $i=0,\ldots, H_{\text{max}}-1$.
The time interval $\delta t$ is taken as $1/D_{\text r}$, which is the average time for a particle A
to jump to a neighboring O site; this mimics a sequential release of particles A at each column.
For each particle A, the transport time is computed as the total number of step trials until
reacting with an M site.
The average time for the thickness to increase from $H=i$ to $H=i+1$ is then calculated as
the average of release time plus transport time in the $i$-th set of particles A.
As the passive layer thickens, the transport times become much larger than the release times,
and consequently the former gives the main contribution to the total growth time.

\section{Growth of the passive layer}
\label{sectiongrowth}

\begin{figure}[!b]
\center
\includegraphics[clip,width=0.95\textwidth, angle=0]{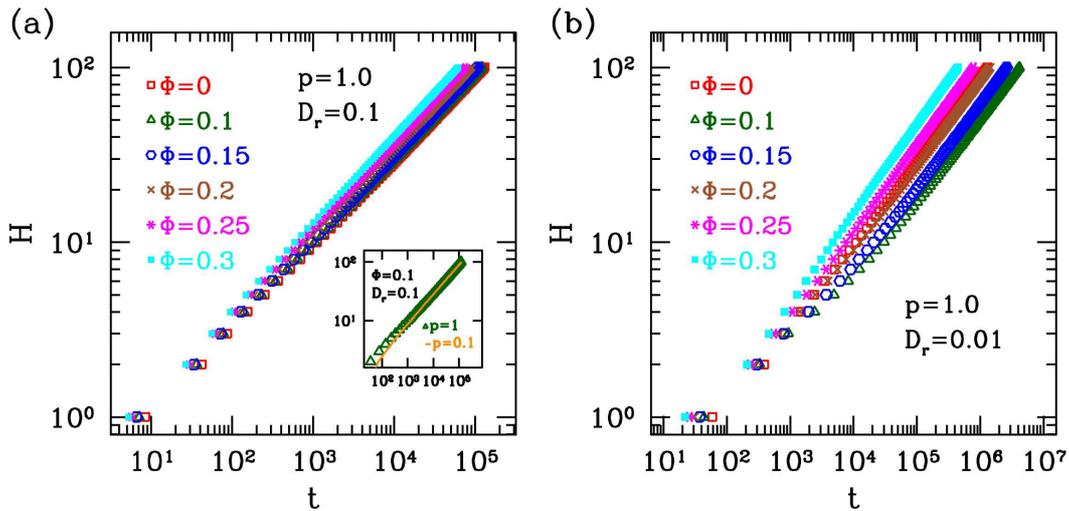}
\caption{(Color online) Thickness of the passive layer as a function of time for the
parameters indicated in the plots.
In (a), the inset compares data for different values of $p$.
}
\label{ht}
\end{figure}

The thickness of the passive layer is shown in figures~\ref{ht}~(a) and \ref{ht}~(b)
as a function of time for $D_{\text r}=0.1$ and $D_{\text r}=0.01$,
respectively, and for several values of the porosity $\Phi$, with
reaction probability $p=1$.
At long times, the results for smaller values of $p$ are similar, as shown in the
inset of figure~\ref{ht}(a).

In all cases, a diffusive displacement of the corrosion front is observed at long times:
\begin{equation}
H \approx D_{\text{CF}} t^{1/2} .
\label{defD}
\end{equation}
Here, $D_{\text{CF}}$ has a dimension of a diffusion coefficient and measures how fast the
corrosion front moves.
Alternatively, a corrosion rate may be defined as the ratio $H/t$, but this is a time
decreasing quantity.
For $\phi=0$, we obtain $D_{\text{CF}}\sim D_{\text r}$; this is the case of a compact layer,
in which our model is equivalent to the long time limit of the model introduced in
 \protect\cite{passive2}; indeed, that model predicts the proportionality between
$D_{\text{CF}}$ and $D_{\text r}$.
As $\Phi$ increases, different trends are observed for $D_{\text r}=0.1$ and $D_{\text r}=0.01$:
in the former, the corrosion rate at a given time increases with $\Phi$, although for
small $\Phi$ this increase is very small; in the latter, a nonmonotonous variation of
that rate is observed, with a decrease for low $\Phi$ and subsequent increase.

\begin{figure}[!t]
\center
\includegraphics[clip,width=0.45\textwidth, angle=0]{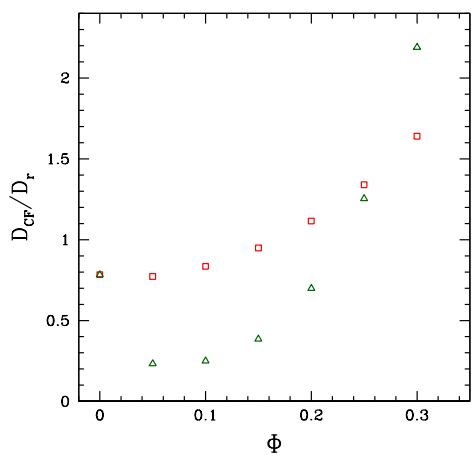}
\caption{(Color online) Estimates of $D_{\text{CF}}/D_{\text r}$ as a function of $\Phi$ for $D_{\text r}=0.1$ (red squares)
and $D_{\text r}=0.01$ (green triangles). In both cases $p=1.0$.
}
\label{amp}
\end{figure}

The estimates of $D_{\text{CF}}/D_{\text r}$ are suitable to characterize those trends.
Figure~\ref{amp} shows $D_{\text{CF}}/D_{\text r}$ as a function of the porosity $\Phi$,
for two values of $D_{\text r}$ and $p=1$.
For $D_{\text r}=0.1$, $D_{\text{CF}}$ has negligible changes for $\Phi\leqslant 0.1$, but increases for large $\Phi$.
The sets of neighboring P sites may form paths where the particles A can move faster, but
isolated P sites have an opposite effect and confine them for long times; when the particle A
reaches an isolated P site, it takes a long time to move to a neighboring O site and there
is a large probability that it returns to the previous P site; see step probabilities in
figure~\ref{model}.
These effects are balanced for small $\Phi$, but an increase in the porosity eventually
leads to an increase of $D_{\text{CF}}$.
The relation between $D_{\text{CF}}$ and $D_{\text r}$ is consequently nonlinear due to the complex
pore connectivity, which is well known from percolation theory \cite{stauffer}.
For $D_{\text r}=0.01$, a small porosity has a much more drastic effect: $D_{\text{CF}}$ decreases by a
factor $\approx 4$ in comparison with the compact oxide layer for $\Phi=5\%$ and $\Phi=10\%$.
This occurs due to the trapping of the particles A in isolated pores for long times.
$D_{\text r}$ begins to increase for larger porosity because longer channels of P sites appear and
compensate that effect.

A diffusive displacement of the corrosion front [equation~(\ref{defD})] was already observed
in the oxidation of iron or iron nitride in different conditions
\cite{caule,sakai,jutte,graat}; however, in some cases it was shown that the
rate limiting process was a diffusion of iron cations \cite{jutte,graat} and was not a
diffusion of anionic species, as proposed in our model.
The diffusive law was also obtained in several models, e.g., those of 
\protect\cite{desgranges2,pena2008,sun2013}.
However, other models and experimental observations suggest a slower growth of
passive layers as a consequence of additional energy barriers for ion diffusion
or dissolution of that layer \cite{chao}.

The results presented here show that porosity of the passive layer has a nontrivial
effect on ion diffusion.
Such effects may also appear in systems with additional energy barriers if the
transport in large pores is much faster than that in the more compact regions of
the passive layer.

\section{Roughness of the corrosion front}
\label{sectionroughness}

The corrosion front may have overhangs because all exposed M sites are subject
to react with particles~A, but to calculate the roughness it is useful to define a
single-valued interface $\left\{ h\right\}$.
At each position $x$ (column $x$), $h$ is the topmost position $z$ of an M site in that column;
the orientation of the $z$ axis is  shown in figure~\ref{model}~(b). The height $h$ is  the minimum value
of $z$ for an M site in that column.

The roughness $W\left( t\right)$ of the corrosion front is defined as the root mean
square fluctuation of the interface $\left\{ h\right\}$ at time $t$.
Here, we analyze the variation of $W$ as a function of the thickness $H$ instead of the
time $t$ because this approach helps the interpretation of the results in the
light of kinetic roughening theory \cite{barabasi,krug}, which is mostly based on models
of interfaces moving with constant velocity.

\begin{figure}[!t]
\centering
\begin{minipage}{0.49\textwidth}
\begin{center}
\includegraphics[width=0.95\textwidth]{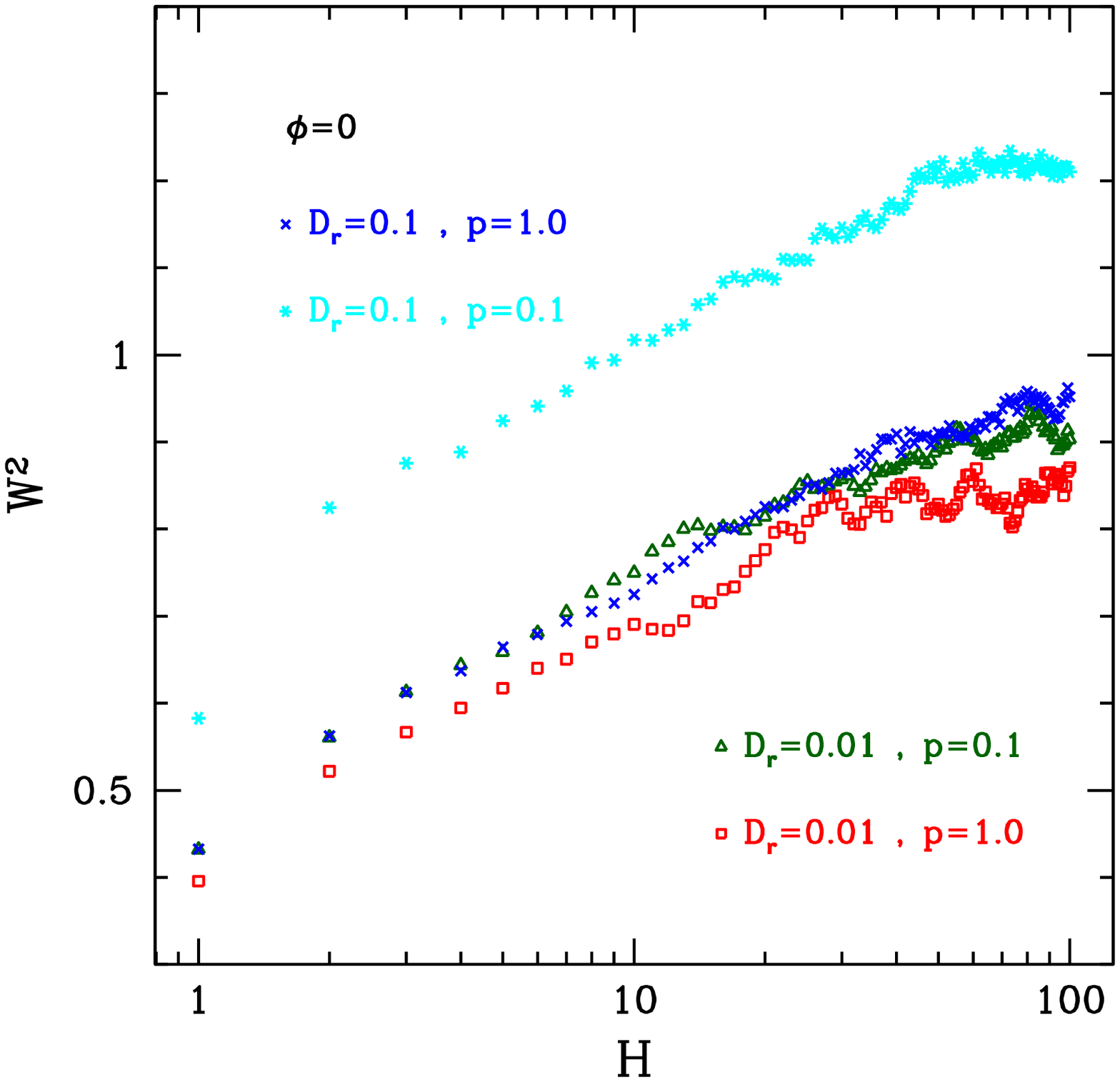}
\caption{(Color online) Roughness of the corrosion front as a function of its average displacement,
for compact layers grown with the values of $D_{\text r}$ and $p$ indicated in the plot.}\label{roughnessP0}
\end{center}
\end{minipage}
\begin{minipage}{0.49\textwidth}
\begin{center}
\includegraphics[width=0.95\textwidth]{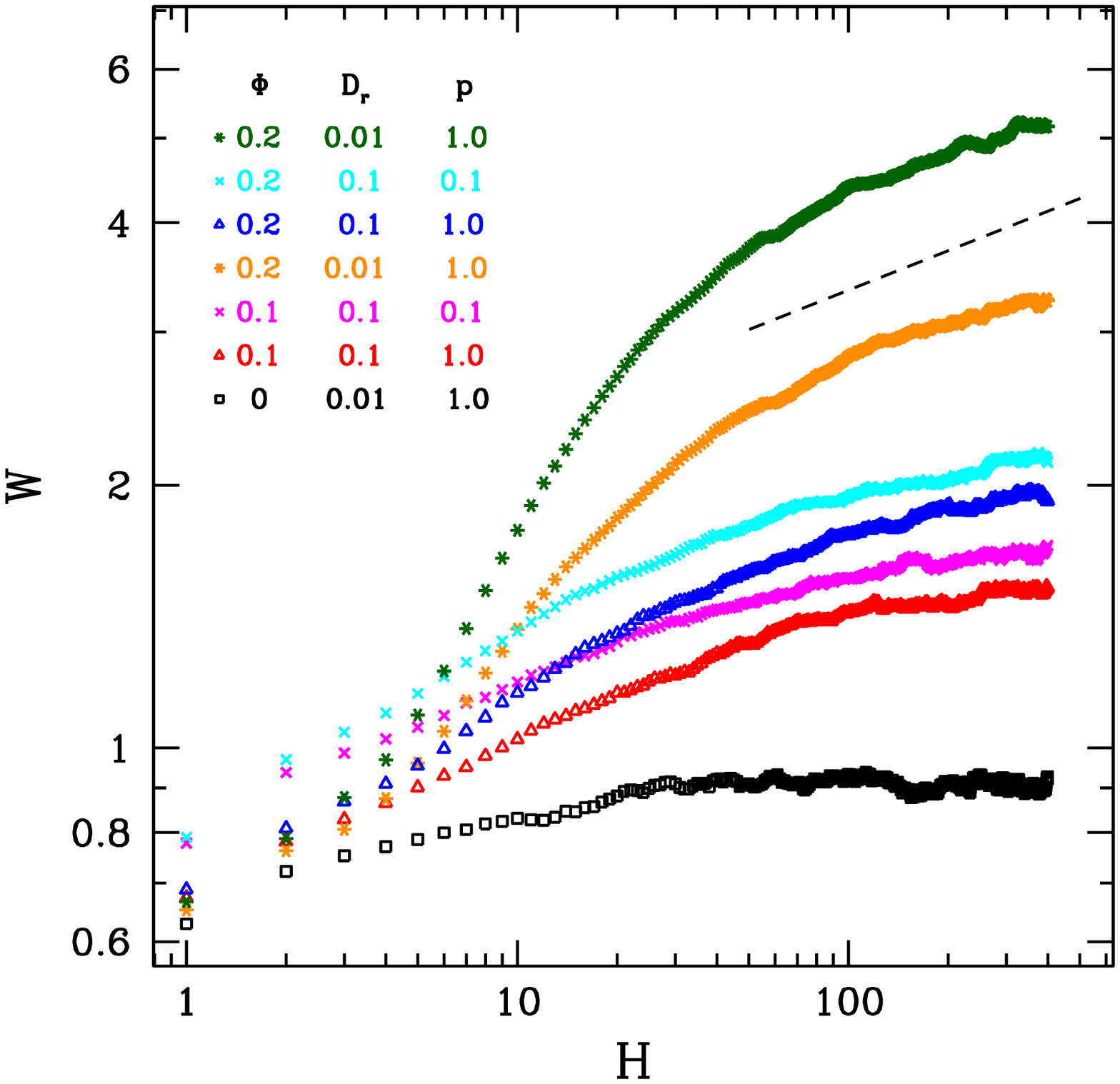}
\caption{(Color online) Roughness of the corrosion front as a function of the passive layer thickness for
the values of the parameters indicated in the plot.
The dashed line has slope $0.15$.}
\label{roughness}
\end{center}
\end{minipage}
\end{figure}

Figure~\ref{roughnessP0} shows the square roughness as a function of $H$ for
compact films ($\Phi =0$), two values of $D_{\text r}$, and two values of $p$.
The semilogarithmic plot was useful to highlight the scaling as
\begin{equation}
W^2 \sim \ln{H}.
\label{WDLE}
\end{equation}
This slow roughening appears because the peaks of the corrosion front (i.e., the
most prominent M sites) have a larger probability of reacting with the incident particles A
than the valleys of that interface.
In these conditions, overhangs in the corrosion front are very rare.

The scaling in equation~(\ref{WDLE}) is explained by the equivalence between our model with
$p=1$ and $\Phi =0$ and the DLE, in which equation~(\ref{WDLE}) is predicted
in two dimensions \cite{krugmeakin}.
Such scaling was also observed in the models of passive layer
growth of \protect\cite{passive1,passive2}.
Figure~\ref{roughnessP0} also shows the saturation of the roughness at short times, which is
expected due to the small value $z=1$ of the dynamical exponent of DLE \cite{krugmeakin}.
In three-dimensions, the DLE theory predicts a finite and small interface roughness at all times,
which is consistent with observations of altered layers in mineral
weathering \cite{alteredlayer}.

Figure~\ref{roughnessP0} shows that the roughness obtained for $D_{\text r}=0.1$ and $p=0.1$ is much
larger than the roughness obtained with other model parameters (although the absolute value
of $W$ is still very small).
This is a case in which the reaction rate is small compared to the rate of diffusion, so that
the particle A scans a large region of the metal-oxide interface before reacting.
In this case,  \protect\cite{passive2} predicts that the roughening at short
times is similar to that of the Eden model \cite{eden,wolf}, in which all exposed M sites
have the same probability to react.
The roughening of the Eden model is in the class of the KPZ equation
\cite{kpz}, which gives $W\sim H^{1/3}$ in two dimensions.
Indeed, for $D_{\text r}=0.1$ and $p=0.1$, an initial rapid roughening is observed in
figure~\ref{roughnessP0} and, for $H \approx 3$, a crossover to the logarithmic scaling of
equation~(\ref{WDLE}) is observed.
Reference \protect\cite{passive2} considered much smaller values of $p$ for compact layers, which
showed crossovers at much larger thicknesses.

In figure~\ref{roughness} we show $W$ as a function of $H$ for the porous films and for
one of the compact deposits.
For $D_{\text r}=0.1$, the presence of porosity leads to an increase in the roughness.
Again, the reduction of the reaction probability $p$ also contributes to the increase of
the roughness.
However, for $D_{\text r}=0.01$, a much larger change in the roughness of porous layers is observed.
For small $D_{\text r}$, the passive layer is highly inhomogeneous to the diffusion of particles A,
which leads to their confinement at the P sites for long times, as discussed above.
This facilitates reactions at the M sites localized in regions with larger local concentrations
of P sites.
The more rapid growth of the front at those regions leads to an increase in the roughness.

Figure~\ref{roughness} shows that the roughness increases approximately as
$W\sim H^{0.15}$ in the thickness range $60\leqslant H\leqslant 400$ for $D_{\text r}=0.01$.
This scaling relation differs from those obtained in the most frequently studied models
of interface growth \cite{barabasi,krug}.
Possibly this occurs because the simulated thicknesses are relatively small and a crossover
to another scaling will appear at longer times.
However, such investigation is beyond the scope of the present work due to the much longer
simulation times that would be required.

\section{Conclusion}
\label{conclusion}

We studied a model of metal corrosion and growth of a passive layer in which an anionic
species diffuses across that layer and reacts at the metal-oxide interface.
The model for a homogeneous layer is similar to that of  \protect\cite{passive2}.
Here, we advanced over that work by considering that large pores may be formed after
the reactions and that the diffusion coefficient of anions in those pores is larger
than that in the compact regions.
The total porosity of such inhomogeneous layers is assumed to be dominated by the
volume fraction of the large pores.

When the ratio between diffusion coefficients in compact regions and in large pores is
small, significant effects of the porosity are observed.
For small porosity, a slowdown of the passive layer growth is observed because anions
are frequently trapped in isolated pores; for the ratio ${10}^{-2}$ considered in our
simulations, this occurs up to $20\%$ of porosity.
On the other hand, the roughness of the corrosion front increases with the porosity
because the random distribution of pores near that front increases the effective noise
amplitude of the roughening process.

The inhomogeneity in our model of passive layer is completely random,
but even in this case it leads to an increase in the roughness of the corrosion front.
This result is consistent with the hypothesis that the defects of an
oxide/hydroxide film formed on a metal are preferential points for nucleation of
metastable pits \cite{amin2014}, i.e., for the initiation of a localized
rapid corrosion which is followed by repassivation.
It is important to recall that other types of random disorder or surface inhomogeneity
in corrosion models are also associated with an increase of roughness of corrosion fronts
\cite{cordoba2007,cordoba2008,constantoudis2009,dung2016}
or with the changes in the morphology of corrosion pits \cite{burridge2015,chen2016,wang2016}.
For these reasons, we believe that an improvement of the present model may be useful
for a quantitative description of the phenomena related to corrosion and passivation.

\section*{Acknowledgements}

This work is dedicated to the memory of my friend and scientific
collaborator Jean-Pierre Badiali, who introduced me to the area of corrosion modelling.

This work was supported by CNPq and Faperj (Brazilian agencies).

\ukrainianpart

\title{Ефекти пористості в моделі корозії та росту пасивного шару}

\author{Ф.Д.А. Аарао Райс}

\address{
Інститут фізики, Федеральний університет Флуміненс, 
Авеню Літоранеа, 24210-340 Нітерой Ріо-де-Жанейро, Бразилія
}

\makeukrtitle

\begin{abstract}
Представлено стохастичну ґраткову модель для вивчення ефектів формування пор у пасивному шарі, який вирощується з продуктами корозії металу. Вважається, що аніони дифундують через цей шар і реагують на корозійній межі (межі розділу металу та оксиду), утворюючи випадковий розподіл компактних областей і великих пор, які відповідають вузлам видів О (оксид) і Р (пора). Припускається, що вузли виду О мають дуже малі пори, тобто частка $\Phi$ вузлів виду P приблизно відповідає пористості, а співвідношення коефіцієнтів дифузії аніонів у цих областях $D_{\text r}<1$.
Результати моделювання для випадку відсутності великих пор ($\Phi =0$) подібні до результатів для попередньої моделі корозії та пасивації і пояснюються масштабним підходом. При $\Phi >0$ і $D_{\text r}\ll 1$ спостерігаються суттєві зміни в рості пасивного шару і жорсткості корозійної межі. 
Для малих $\Phi$ зауважено сповільнення швидкості росту, яке можна пояснити тривалим обмеженням руху аніонів в ізольованих порах.
Однак присутність великих пор поблизу корозійної межі збільшує частоту реакцій в цих областях, приводячи до збільшення жорсткості межі. Дану модель можна розглядати як перший крок при описі дефектів у пасивному шарі, які сприяють герметичній корозії.

\keywords стохастична модель, корозія, пасивація, дифузія, пористість, жорсткість

\end{abstract}

\end{document}